\documentclass[twocolumn,showpacs,prb,superscriptaddress]{revtex4}

\usepackage{graphicx}%
\usepackage{dcolumn}
\usepackage{amsmath}

\usepackage{color}

\makeatletter
\def\btt#1{\texttt{\@backslashchar#1}}%
\DeclareRobustCommand\bblash{\btt{\@backslashchar}}%
\makeatother

\begin{document}

\preprint{HEP/123-qed}

\title[Short Title]{Electrically driven spin excitation in a ferroelectric magnet DyMnO$_{\textbf 3}$}

\author{N. Kida}
\affiliation{Multiferroics Project (MF), ERATO, Japan Science and Technology Agency (JST), c/o Department of Applied Physics, The University of Tokyo, 7-3-1 Hongo, Bunkyo-ku, Tokyo 113-8656, Japan}

\author{Y. Ikebe}
\affiliation{Department of Physics, The University of Tokyo, 7-3-1 Hongo, Bunkyo-ku, Tokyo 113-0033, Japan}

\author{Y. Takahashi}
\affiliation{Multiferroics Project (MF), ERATO, Japan Science and Technology Agency (JST), c/o Department of Applied Physics, The University of Tokyo, 7-3-1 Hongo, Bunkyo-ku, Tokyo 113-8656, Japan}

\author{J. P. He}
\affiliation{Multiferroics Project (MF), ERATO, Japan Science and Technology Agency (JST), c/o Department of Applied Physics, The University of Tokyo, 7-3-1 Hongo, Bunkyo-ku, Tokyo 113-8656, Japan}

\author{Y. Kaneko}
\affiliation{Multiferroics Project (MF), ERATO, Japan Science and Technology Agency (JST), c/o Department of Applied Physics, The University of Tokyo, 7-3-1 Hongo, Bunkyo-ku, Tokyo 113-8656, Japan}

\author{Y. Yamasaki}
\affiliation{Department of Applied Physics, The University of Tokyo, 7-3-1 Hongo, Bunkyo-ku, Tokyo 113-8656, Japan}

\author{R. Shimano}
\affiliation{Multiferroics Project (MF), ERATO, Japan Science and Technology Agency (JST), c/o Department of Applied Physics, The University of Tokyo, 7-3-1 Hongo, Bunkyo-ku, Tokyo 113-8656, Japan}
\affiliation{Department of Physics, The University of Tokyo, 7-3-1 Hongo, Bunkyo-ku, Tokyo 113-0033, Japan}

\author{T. Arima}
\affiliation{Institute of Multidisciplinary Research for Advanced Materials, Tohoku University, 2-1-1 Katahira, Aoba-ku, Sendai 980-8577, Japan}

\author{N. Nagaosa}
\affiliation{Department of Applied Physics, The University of Tokyo, 7-3-1 Hongo, Bunkyo-ku, Tokyo 113-8656, Japan}
\affiliation{Cross-Correlated Materials Research Group (CMRG), ASI, RIKEN, 2-1 Hirosawa, Wako, 351-0198, Japan}

\author{Y. Tokura}
\affiliation{Multiferroics Project (MF), ERATO, Japan Science and Technology Agency (JST), c/o Department of Applied Physics, The University of Tokyo, 7-3-1 Hongo, Bunkyo-ku, Tokyo 113-8656, Japan}
\affiliation{Department of Applied Physics, The University of Tokyo, 7-3-1 Hongo, Bunkyo-ku, Tokyo 113-8656, Japan}
\affiliation{Cross-Correlated Materials Research Group (CMRG), ASI, RIKEN, 2-1 Hirosawa, Wako, 351-0198, Japan}

\date{\today}

\begin{abstract}
Temperature (5--250 K) and magnetic field (0--70 kOe) variations of the low-energy (1--10 meV) electrodynamics of spin excitations have been investigated for a complete set of light-polarization configurations for a ferroelectric magnet DyMnO$_3$ by using terahertz time-domain spectroscopy. We identify the pronounced absorption continuum (1--8 meV) with a peak feature around 2 meV, which is electric-dipole active only for the light $E$-vector along the $a$-axis. This absorption band grows in intensity with lowering temperature from the spin-collinear paraelectric phase above the ferroelectric transition, but is independent of the orientation of spiral spin plane ($bc$ or $ab$), as shown on the original $P_{\rm s}$ (ferroelectric polarization) $\parallel c$ phase as well as the magnetic field induced $P_{\rm s}\parallel a$ phase. The possible origin of this electric-dipole active band is argued in terms of the large fluctuations of spins and spin-current.
\end{abstract}


\maketitle

%***********************************************************************
\section{Introduction}
The magnetoelectric (ME) effect, in which the electric polarization can be induced by a magnetic field $H$ or the magnetization can be induced by an electric field $E$, has been known since the 1950's.\cite{MEeffectRev} However, the materials showing the large ME effect are rare. Recently, the interest has been revived by the observation of the ferroelectric polarization $P_{\rm s}$ and its 90$^\circ$ flop in the presence of a finite field $H$,\cite{TKimura1} in orthorhombically distorted perovskite manganites, $R$MnO$_3$, where $R$ represents Tb, Dy, and Gd ions.\cite{TKimura1,TGoto,TKimura2} In DyMnO$_3$, for example, the collinear spin order of Mn ions sets in at the N\'{e}el temperature $T_{\rm N}^{\rm Mn}$ of 39 K with the incommensurate (IC) modulation vector $q_b^{\rm Mn}\sim0.36$ along the $b$-axis.\cite{TGoto} Noticeably, the ferroelectricity shows up along the $c$-axis when the transverse $bc$ spiral (cycloidal) spin order\cite{MKenzelmann} evolves below $T_{\rm c}$ of 19 K. The direction of $P_{\rm s}$ is flopped to the $a$-axis when $H$ is applied along the $b$-axis.\cite{TGoto,TKimura2} So far, a wide variety of the ferroelectric magnets, termed multiferroics, have been found in a family of cycloidal-spin magnets.\cite{MEeffectRev} The ferroelectricity in $R$MnO$_3$ can be microscopically explained by the spin-current model\cite{HKatsura1} or equivalently by the inverse Dzyaloshinski-Moriya effect,\cite{IASergienko,MMostovoy} as expressed by the relation, $P_{\rm s}\propto e_{ij}\times (S_i\times S_j)$, where $e_{ij}$ is the unit vector connecting $i$ and $j$ sites. Namely, the cycloidal spin structure can induce the ferroelectricity along the direction perpendicular to the modulation vector and within the spiral spin plane. A direct evidence for this mechanism was recently demonstrated by controlling the vector chirality ($S_i \times S_j$) of TbMnO$_3$ with $E$.\cite{YYamasaki}

In the magnetically induced ferroelectric phase, the strong ME coupling would also produce the intriguing low-lying optical excitation. In $R$MnO$_3$, there are experimental (Ref. \onlinecite{APimenov1}) and theoretical\cite{GASmolenski,HKatsura2} arguments on the possibility of the presence of the low-lying spin excitation, called {\it electromagnon}, which is activated electrically. Far-infrared spectroscopy has provided the useful information for the electrodynamics of TbMnO$_3$,\cite{APimenov1} GdMnO$_3$,\cite{APimenov1,APimenov2} and Eu$_{1-x}$Y$_x$MnO$_3$,\cite{RValdesAguilar,APimenov3} revealing that an additional light absorption contribution indeed emerges in the ferroelectric phase at terahertz (THz) frequencies below the optical phonon frequencies of the perovskite structure. Among a family of $R$MnO$_3$, DyMnO$_3$ shows the largest ME response; the change of the dielectric constant upon the $P_{\rm s}$ flop exceeds 500\% and $P_{\rm s}$ reaches 2 mC/m$^2$.\cite{TGoto,TKimura2} Therefore, DyMnO$_3$ offers a good opportunity to clarify the low-energy electrodynamics coupled with the spin excitation. 

Here we report on a complete set of optical spectra at THz frequencies of a multiferroic DyMnO$_3$ with varying the light-polarization (both the electric field $E^\omega$ and magnetic field $H^\omega$ components of light), temperature $T$, and applied $H$, to unravel the strongly electric-dipole allowed spin excitation. The observed excitation is confirmed to be active mainly for $E^\omega \parallel a$ and not to be correlated to the direction of the spiral spin plane. We focus on spectral signatures, being highly sensitive to the light-polarization, in a variety of ME phases that can be tuned by both $T$ and $H$. The format of this paper is as follows. In Sec. \ref{Methods} we describe the detail of the experimental procedure for measurements of THz time-domain spectroscopy. Section \ref{Results} is devoted to the optical investigations of the low-energy spin dynamics in a variety of ME phases of DyMnO$_3$, in which we clarify the unique selection rule of the spin excitation driven by $E^\omega$ or $H^\omega$ by measurements of the light-polarization (Sec. \ref{ResultsLP}), $T$ (Sec. \ref{ResultsT}), and $H$ (Sec. \ref{ResultsH}) dependences.  On the basis of the results of Sec. \ref{Results}, the origin of the electrically driven spin excitation observed at THz frequencies in DyMnO$_3$ is discussed in Sec. \ref{Discussion}. The summary of this paper is given in Sec. \ref{Summary}. We also describe in Appendixes the estimation procedure of the complex optical constants from the raw data obtained by THz time-domain spectroscopy.

%***********************************************************************
\section{Methods}\label{Methods}
%***********************************************************************
\subsection{Sample preparation and characterization}

The large single crystals of DyMnO$_3$ were grown by a floating zone method. The available $ac$, $ab$, and $bc$ surfaces of the crystals with a typical sample size of 4--16 mm$^2$ were cut from the ingot by using a back Laue photograph. The respective samples were polished to thickness of 150--800 $\mu$m. The polishing procedure frequently induces the damage on the surface for the case of the orthorhombically distorted perovskite manganites. We confirmed that the optical response was almost unchanged before and after polishing procedure. 

The samples cut from the same ingot used in this work were characterized by the measurements of the dielectric constant at low-frequencies as functions of $T$ and $H$. The ferroelectricity along the $c$-axis in zero $H$ and the $P_{\rm s}$ flop from the $c$ to $a$-axes in $H$ were clearly observed around 19 K and 20 kOe at 7 K, respectively. The obtained phase diagrams in a plane of $T$ and $H$ are consistent with the previously reported one of DyMnO$_3$.\cite{TKimura2}

%******************************************************** Figure 1
\begin{figure}[bt]
\includegraphics[width=0.48\textwidth]{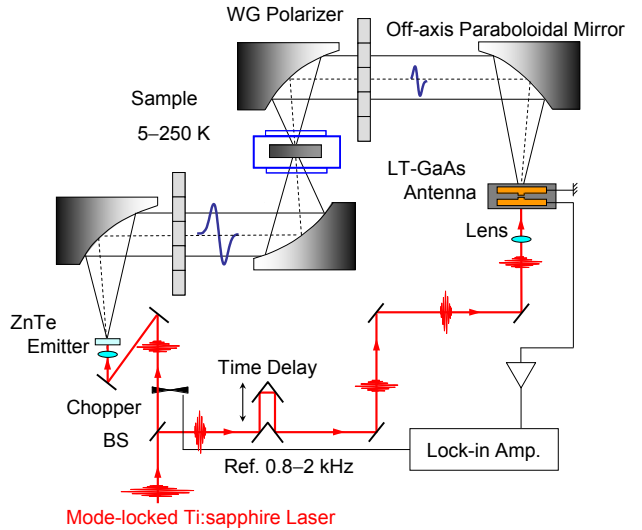}
\caption{(Color online) Schematic illustration of the experimental setup for THz time-domain spectroscopy in a transmission geometry.}
\label{fig1}
\end{figure}
%********************************************************

%***********************************************************************
\subsection{Setup for terahertz time-domain spectroscopy}

We used the THz time-domain spectroscopy in a transmission geometry to obtain complex refractive index $\tilde{n}$ without using Kramers-Kronig (KK) analysis.\cite{THzRev,MTonouchiRev} THz time-domain spectroscopy can easily access the low-energy electrodynamics in solids, typical frequencies from 0.1 THz to 3 THz.\cite{THzRev} The lower limit of the measurable frequencies depends on the size of the samples since 0.1 THz corresponds to 3 mm in wavelength. We adopted a transmission geometry and the sample was attached to the perforated Cu holder with a diameter of 2 mm, 2.8 mm, 3 mm, and 3.5 mm, which was selected to suitably match the size of the samples. 

Figure \ref{fig1} shows the schematic illustration of THz time-domain spectroscopy in a transmission geometry in this study. For the measurements, we used the photoconducting (PC) sampling technique\cite{THzRev1} to obtain THz radiation pulse. The femtosecond laser pulses delivered from the mode-locked Ti:sapphire laser (center wavelength of 800 nm, pulse width of 90 fs, and repetition rate of 80 MHz) were used as a source. The laser pulses were divided by a beam splitter (BS). One was used as the pump and another was used as the gate. The pump pulses were irradiated on to the ZnTe crystal\cite{QWu} to induce the optical rectification effect, acting as a source of THz radiation.\cite{KYang} The wire grid (WG) polarizers were inserted in between off-axis paraboloidal mirrors to obtain the linear polarization. The polarized THz pulse was focused on the sample by off-axis paraboloidal mirrors. The cryostat was placed within the box filled with dry nitrogen to eliminate the absorption of water. The gate pulses were introduced to the low-temperature-grown GaAs (LT-GaAs) photoconducting device coupled with a dipole antenna after the appropriate time delay.\cite{DHAuston} The pump pulses were mechanically chopped and the photocurrent induced by the electric field of the THz pulse was lock-in detected. Therefore, the induced photocurrent, which depends on both amplitude and phase of THz radiation, can be obtained by varying the optical delay line. Further details of our scheme for THz generation and detection can be found in Refs. \onlinecite{THzRev} and \onlinecite{THzRev1}.

For the measurements of the $H$ dependence, the sample was placed in a cryostat equipped with a superconducting magnet. $H$ supplied from a superconducting magnet was applied along the $b$-axis up to 70 kOe. The measurements in $H$ were mainly performed at 7 K. We adopted a Faraday geometry. Further details of our experimental setup combined with the superconducting magnet can be found in Ref. \onlinecite{YIkebe}.

%******************************************************** Figure 2
\begin{figure*}[bt]
\includegraphics[width=0.8\textwidth]{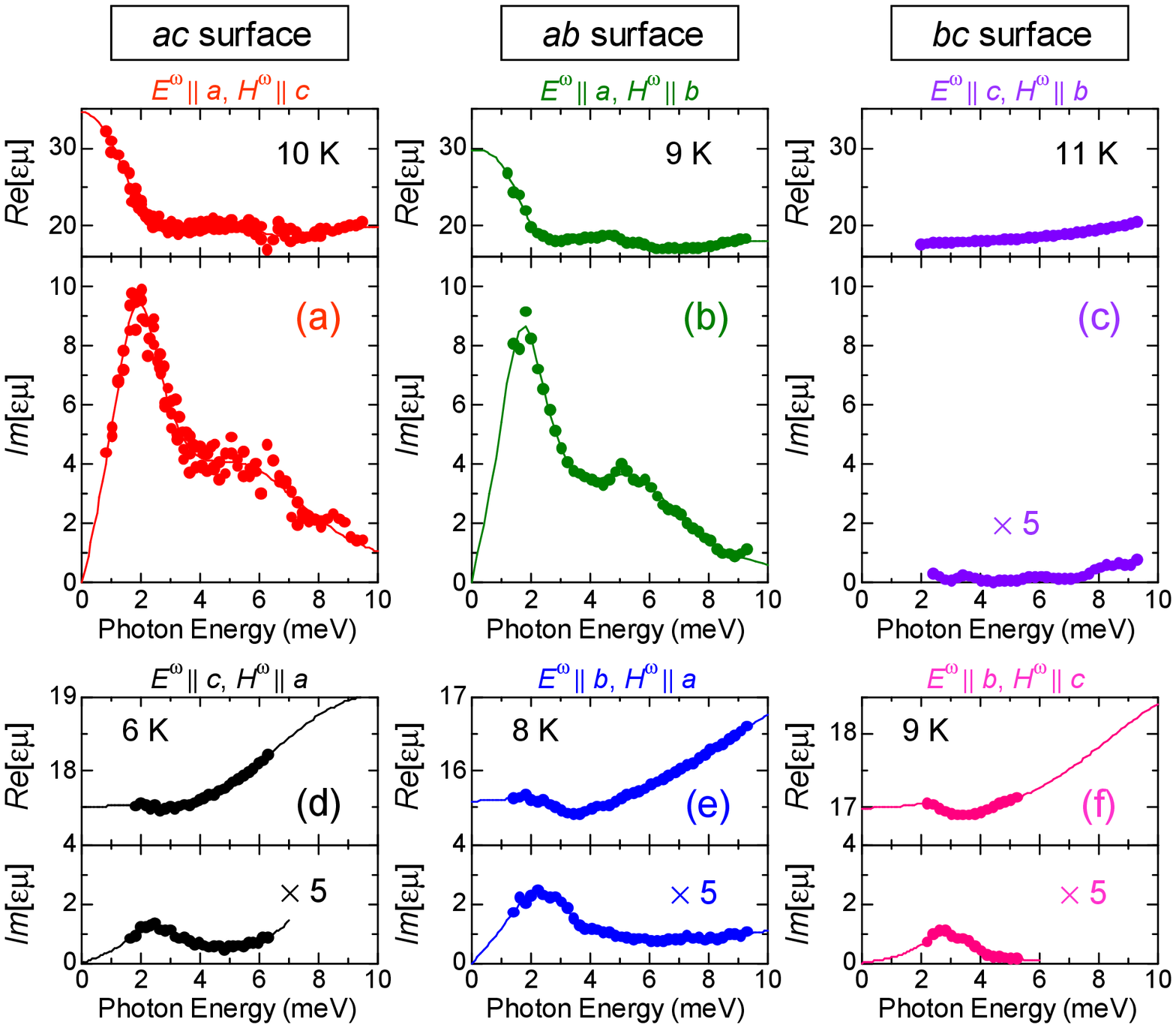}
\caption{(Color online) Low-energy electrodynamics of a complete set of available configurations of single crystals of DyMnO$_3$, measured around 10 K in zero magnetic field. Upper and lower panels show the real and imaginary parts of $\epsilon\mu$ spectrum (closed circles) when $E^\omega$ and $H^\omega$ were set parallel to the crystallographic axes of $ac$, $ab$, and $bc$ surface crystal plates. The solid lines in (a) and (b) are results of a least-squares fit to reproduce low- and high-lying peak structures by assuming two Lorentz oscillators for $\epsilon$. On the contrary, $\epsilon\mu$ spectra in (d), (e), and (f) can be reproduced by two Lorentz oscillators for $\epsilon$ and $\mu$, as indicated by solid lines.}
\label{fig2}
\end{figure*}
%******************************************************** 

%******************************************************** Figure 3
\begin{figure*}[bt]
\includegraphics[width=0.95\textwidth]{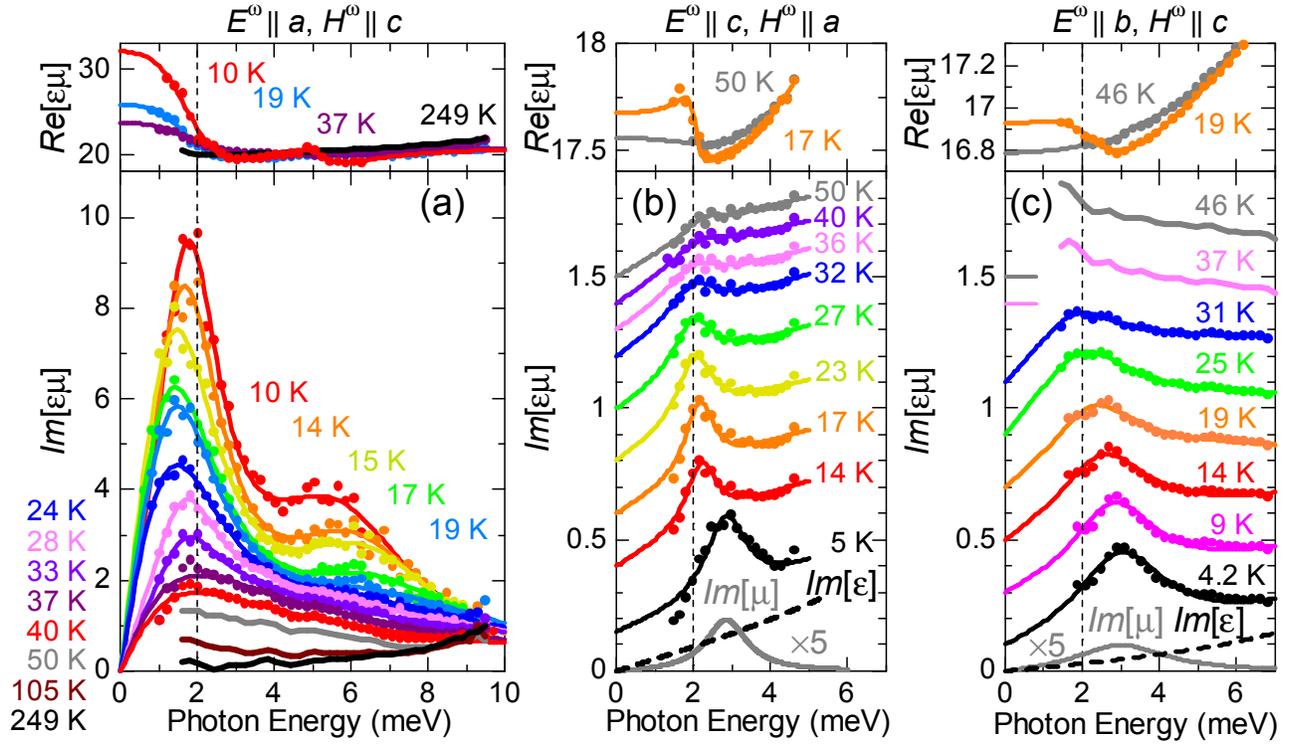}
\caption{(Color online) Temperature dependence of the real (upper panels) and imaginary (lower panels) parts of the selected $\epsilon\mu$ spectra (closed circles), of DyMnO$_3$ for (a) $E^\omega\parallel a$, (b) $H^\omega\parallel a$, and (c) $H^\omega\parallel c$. Note that the scale of the vertical axis is different in respective figures. The $Im[\epsilon\mu]$ spectra shown in (b) and (c) are vertically offset for clarity. We also plot the extracted $Im[\epsilon]$ and $Im[\mu]$ spectra [at (b) 5 K and (c) 4.2 K] indicated by dotted and gray lines, respectively, by assuming that $Im[\epsilon\mu]$ spectrum can be represented by two Lorentz oscillators for $\epsilon$ and $\mu$ (solid lines). The $Im[\mu]$ spectrum is multiplied by 5. On the contrary, $\epsilon\mu$ spectra for $E^\omega\parallel a$ [(a)] can be fitted by two Lorentz oscillators.}
\label{fig3}
\end{figure*}
%********************************************************

Since we consider the spin excitation, which can be driven by either $E^\omega$ or $H^\omega$, we numerically estimated $\tilde{n}$ and used the quantity of $\tilde{n}^2=\epsilon\mu(\omega)$ (not $\tilde{n}^2=\epsilon$), where $\epsilon$ is the complex dielectric constant and $\mu$ the complex magnetic permeability. Details of the estimation procedure and the validity of the quantities of $\epsilon\mu$ to characterize the spin excitation are described in Appendixes.

%***********************************************************************
\section{Results}\label{Results}
%***********************************************************************
\subsection{Light-polarization dependence}\label{ResultsLP}

First, in Fig. \ref{fig2}, we show $\epsilon\mu$ spectra (closed circles) in the $bc$ spiral spin phase ($P_{\rm s} \parallel c$) of DyMnO$_3$ to determine the selection rule with respect to the light-polarization, in which $E^\omega$ and $H^\omega$ were set parallel to the crystallographic axes using available $ac$, $ab$, and $bc$ surfaces of the crystal. Upper and lower panels of Fig. \ref{fig2} present, respectively, the real $(Re)$ and imaginary $(Im)$ parts of $\epsilon\mu$ spectra, measured around 10 K with $H=0$. There are remarkable peak structures and noticeable optical anisotropy with respect to $E^\omega$ and $H^\omega$. Such a light-polarization dependence, including the negligible absorption for $E^\omega \parallel c$ and $H^\omega \parallel b$ [Fig. \ref{fig2}(c)], as well as the conspicuous spectral change around $T_{\rm c}$ and $T_{\rm N}^{\rm Mn}$, cannot be ascribed to the crystalline-electric-field excitation of $f$ electrons, which was reported for TbMnO$_3$ around 4.7 meV by inelastic neutron scattering studies.\cite{RKajimoto,DSenff}

A broad and pronounced absorption continuum (1--8 meV) overlapped by two peak structures around 2 meV and 5.5 meV is observed in $Im[\epsilon\mu]$ spectrum for $E^\omega \parallel a$ and $H^\omega \parallel c$ [lower panel of Fig. \ref{fig2}(a)]. In accord with this, a dispersive structure can be discerned in $Re[\epsilon\mu$] spectrum [upper panel of Fig. \ref{fig2}(a)]. Such a low-lying peak structure was also identified previously at 2.9 meV for TbMnO$_3$,\cite{APimenov1} 2.5 meV for GdMnO$_3$,\cite{APimenov1,APimenov2} and 3 meV for Eu$_{1-x}$Y$_x$MnO$_3$ (although this value depends slightly on $x$).\cite{RValdesAguilar,APimenov3} Noticeably, $Im[\epsilon\mu]$ of DyMnO$_3$ reaches 10 (corresponding to the optical conductivity of about 3 $\Omega^{-1}$ cm$^{-1}$), which is roughly a factor of 2--5 larger than the absorption of TbMnO$_3$ and GdMnO$_3$, suggesting DyMnO$_3$ as the prototype in the study of the low-lying excitation. Such a strong absorption including the spectral shape, peak position, and magnitude, was clearly confirmed to be observed only for $E^\omega \parallel a$, as evidenced by the $\epsilon\mu$ spectra for $E^\omega \parallel a$ and $H^\omega \parallel b$  [Fig. \ref{fig2}(b)], i.e., with different $H^\omega$ polarization. The low-temperature spectrum can be fitted with two Lorentz oscillators, as indicated by solid lines. The ratio of $Re[\epsilon\mu]$ for $E^\omega \parallel a$ to other configurations was about 2, being consistent with the dielectric constant $Re[\epsilon]$ at 10 kHz.\cite{TGoto,TKimura2} This indicates the large anisotropy extending up to THz frequencies, associated with the electric-dipole active excitation.

In $Im[\epsilon\mu]$ spectrum for $E^\omega \parallel c$ and $H^\omega \parallel a$ [Fig. \ref{fig2}(d)], on the other hand, a broad peak feature is observed where the peak energy nearly matches the low-lying peak energy for $E^\omega \parallel a$. However, the magnitude of $Im[\epsilon\mu]$ in this configuration ($\sim0.3$) is much smaller, by a factor of 30, than the magnitude of $Im[\epsilon\mu]$ for $E^\omega \parallel a$. By comparing the $\epsilon\mu$ spectra for $E^\omega \parallel c$ and $H^\omega \parallel a$ [Fig. \ref{fig2}(d)] and $E^\omega \parallel b$ and $H^\omega \parallel a$ [Fig. \ref{fig2}(e)], this absorption appears to arise from the $H^\omega \parallel a$ stimulation, in accord with the conventional magnon selection rule for the $bc$ spiral spin state.\cite{HKatsura2} We also discern the broad peak feature in $\epsilon\mu$ spectrum for $E^\omega \parallel b$ and $H^\omega \parallel c$ [Fig. \ref{fig2}(f)] whose magnitude of $Im[\epsilon\mu]$ $(\sim0.3)$ is comparable to those of $Im[\epsilon\mu]$ $(\sim0.3-0.5)$ for cases of $H^\omega\parallel a$ stimulation. We cannot distinguish the selection rule of this tiny absorption on the basis of the light-polarization dependence alone due to the presence of the intense absorption for $E^\omega\parallel a$ and $H^\omega\parallel c$ [Fig. \ref{fig2}(a)]. As will be shown in Sec. \ref{ResultsT}, however, we identify that this sharp peak structure is also driven by $H^\omega$.

%******************************************************** Figure 4
\begin{figure}[tbh]
\includegraphics[width=0.35\textwidth]{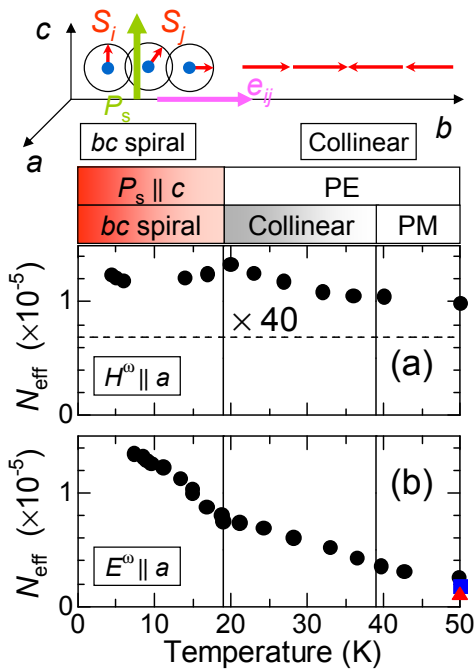}
\caption{(Color online) Temperature $(T)$ dependence of the integrated spectral weight per Mn-site, $N_{\rm eff}$, as defined by Eq. (\ref{eqn.Neff}) in the text, of the $ac$ surface crystal plate of DyMnO$_3$ for $H^\omega \parallel a$ [(a)] and $E^\omega \parallel a$ [(b)]. The upper panel of (a) shows the phase diagram with variation of $T$. PE and PM stand for paraelectric and paramagnetic, respectively. We also show the schematic illustrations of the $bc$ spiral and collinear spin states induced by $T$. N\'{e}el temperature $T_{\rm N}^{\rm Mn}$ (39 K) and ferroelectric transition temperature $T_{\rm c}$ (19 K) are indicated by vertical solid lines in (a) and (b). The data measured at 75 K and 249 K were also plotted at the 50 K position in (b) with square and triangle symbols, respectively. The horizontal dashed line in (a) represents the estimated contribution of the optical phonon absorption at 5 K.}
\label{fig4}
\end{figure}
%********************************************************

%******************************************************** Figure 5
\begin{figure*}[bt]
\includegraphics[width=0.78\textwidth]{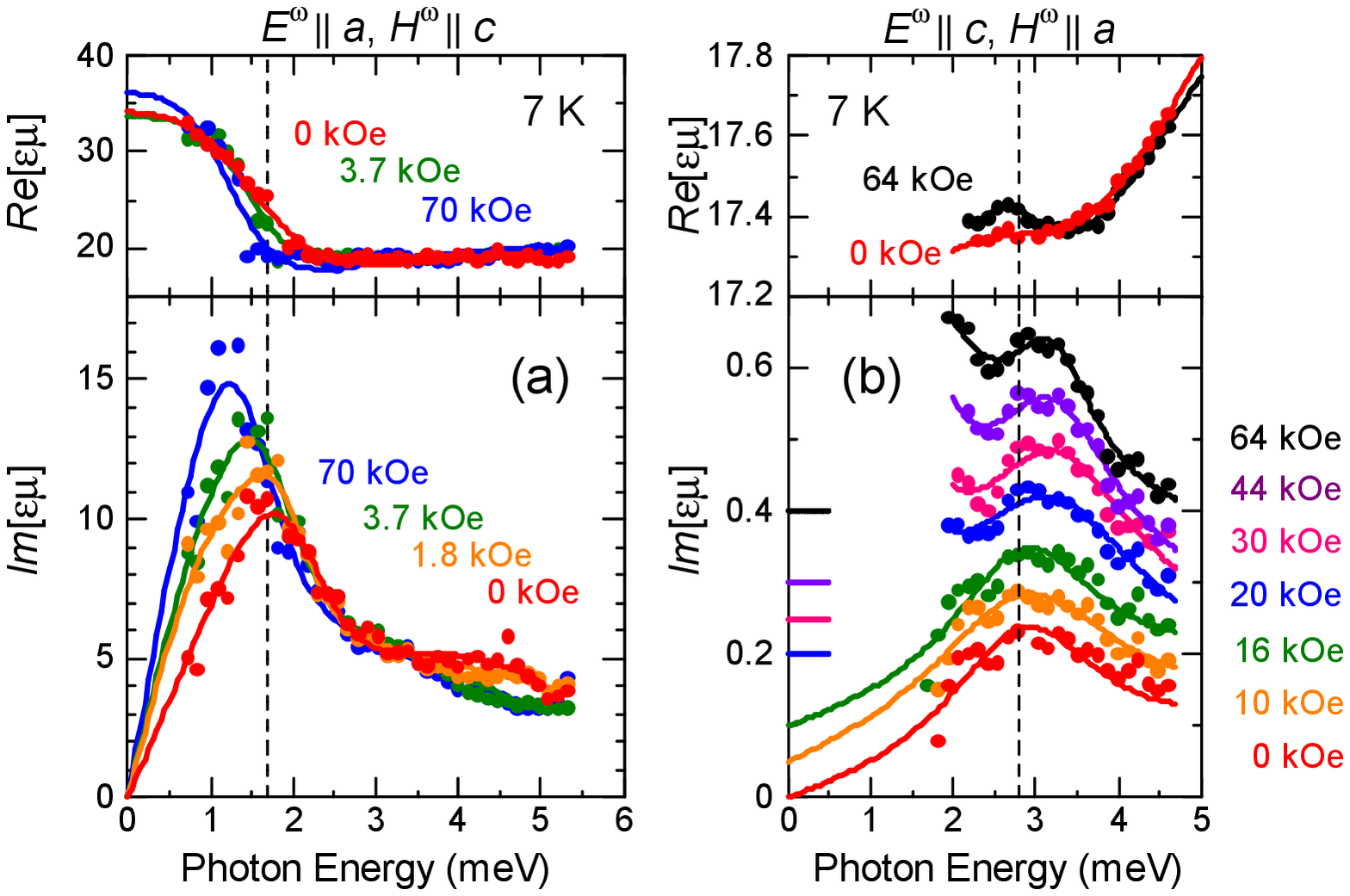}
\begin{minipage}{3.7cm}
\vspace*{-6.9cm}
\caption{(Color online) Magnetic field $(H)$ dependence of the real (upper panels) and imaginary (lower panels) parts of the selected $\epsilon\mu$ spectra (closed circles) of the $ac$ surface crystal plate of DyMnO$_3$ at 7 K when (a) $E^\omega$ or (b) $H^\omega$ was set parallel to the $a$-axis. The ferroelectric polarization flop occurs at 20 kOe for $H \parallel b$. Note that the $Im[\epsilon\mu]$ spectra in (b) are vertically offset for clarity and that the scale of the vertical axis is different in (a) and (b).}
\label{fig5}
\end{minipage}
\end{figure*}
%********************************************************

%***********************************************************************
\subsection{Temperature dependence}
\label{ResultsT}

We present in Fig. \ref{fig3}(b) the $T$ dependence of the selected $\epsilon\mu$ spectra for $H^\omega \parallel a$. With increasing $T$, $Im[\epsilon\mu]$ tends to decrease in magnitude and the peak position is shifted to the lower energy from 2.8 meV at 5 K to 2 meV at 36 K [the lower panel of Fig. \ref{fig3}(b)]. Above $T_{\rm N}^{\rm Mn}$, the broad peak structure is diminished, as can be seen in $Im[\epsilon\mu]$ spectrum at 50 K (a finite value of $Im[\epsilon\mu]$ comes from the absorption tail of the optical phonons). Accordingly, there is no dispersive structure around 2 meV in $Re[\epsilon\mu]$ spectrum [upper panel of Fig. \ref{fig3}(b)]. We also show in Fig. \ref{fig3}(c) the selected $\epsilon\mu$ spectra for $H^\omega\parallel c$ as a function of $T$. With increasing $T$, the peak position is also shifted to the lower energy from 3.3 meV at 4.2 K to 2.3 meV at 31 K and the magnitude of $Im[\epsilon\mu]$ also tends to decrease. The observed $T$ variation is similar to the case for $H^\omega\parallel a$; the peak structure seems to vanish above $T_{\rm N}^{\rm Mn}=39$ K. Even though we cannot identify the clear signature of the peak structure above 31 K in $Im[\epsilon\mu]$ spectrum [the lower panel of Fig. \ref{fig3}(c)] due to the detection limit in our experiments ($\sim1.8$ meV), as determined by the size of the sample, there is no signature of the dispersive structure around 2 meV in $Re[\epsilon\mu]$ spectrum at 46 K [the upper panel of Fig. \ref{fig3}(c)]. On the basis of the $T$ dependence as well as of the light-polarization dependence presented in Sec. \ref{ResultsLP}, it is reasonable to consider that the observed peak structure can be assigned to the spin excitation of Mn ions driven by $H^\omega \parallel a$ or $H^\omega\parallel c$. Indeed, in inelastic neutron scattering experiments on the $bc$ spiral spin phase of TbMnO$_3$, the magnon dispersion along the [100] or [001] axis comes across $k=0$ around 2 meV.\cite{RKajimoto,DSenff} To estimate the contribution of the magnon to $Im[\epsilon\mu]$ spectrum, we plot in Figs. \ref{fig3}(b) and \ref{fig3}(c) the $Im[\mu]$ spectra for $H^\omega\parallel a$ at 5 K and for $H^\omega\parallel c$ at 4.2 K, respectively, as indicated by solid lines. These were derived from the relation, $Im[\epsilon\mu]$ $(=Re[\epsilon]Im[\mu]+Im[\epsilon]Re[\mu])$ by assuming Lorentz oscillators of $\epsilon$ and $\mu$ for contributions of phonon (a dotted line) and magnon, respectively. 

Next, we show in Fig. \ref{fig3}(a) the $T$ dependence of the $\epsilon\mu$ spectra for $E^\omega \parallel a$. At 249 K, only the absorption tail due to the optical phonon is observed. With decreasing $T$ but even above $T_{\rm N}^{\rm Mn}$, it is clearly discernible that $Im[\epsilon\mu]$ in the low-energy region below 8 meV increases and forms a Debye-like absorption band [the lower panel of Fig. \ref{fig3}(a)]. In accord with this, the dispersive structure emerges in $Re[\epsilon\mu]$ spectrum [the upper panel of Fig. \ref{fig3}(a)]. In the collinear phase below $T_{\rm N}^{\rm Mn}$ (= 39 K), a broad peak structure emerges around 2 meV and grows in intensity continuously with further decreasing $T$. In the $bc$ spiral spin phase below $T_{\rm c}$ (=19 K), the additional peak structure shows up around 6 meV and its peak position shifts to the lower energy while the low-lying peak structure shifts to higher energy.

To be more quantitative, we deduced the integrated spectral weight per Mn-site, $N_{\rm eff}$, as given by
\begin{equation}
N_{\rm eff}=\frac{2m_0V}{\pi e^2}\int_{\omega_1}^{\omega_2}\omega' Im[\epsilon(\omega ')\mu(\omega ')]d\omega ',
\label{eqn.Neff}
\end{equation}
where $m_0$ is the free electron mass, $e$ the elementary charge, and $V$ the formula (DyMnO$_3$) cell volume. The integrated energy range was chosen between $\omega_1$ = 0.8 meV and $\omega_2$ = 9.5 meV to nearly cover the full band. Figure \ref{fig4}(b) presents $T$ variation of the estimated $N_{\rm eff}$. We also plotted $N_{\rm eff}$ at 75 K (square) and 245 K (triangle) at the 50 K position for clarity. $N_{\rm eff}$ for $E^\omega \parallel a$ gradually increases below $T_{\rm N}^{\rm Mn}=39$ K and is sharply enhanced below $T_{\rm c}=19$ K (indicated by vertical solid lines). This is in contrast to $N_{\rm eff}$ for $H^\omega \parallel a$ (we adopted $\omega_1$ = 1.5 meV and $\omega_2$ = 4.6 meV to estimate $N_{\rm eff}$ of the magnon band), as shown in Fig. \ref{fig4}(a); $N_{\rm eff}$ reaches the maximum at $T_{\rm c}$ where the $bc$ spiral spin structure evolves from the collinear spin state [note that $N_{\rm eff}$ for $H^\omega \parallel a$ includes the contribution of the phonon structure, as indicated by a dashed line in Fig. \ref{fig4}(a), which could be crudely estimated with the $Im[\epsilon]$ spectrum at 5 K, as indicated by a dotted line in Fig. \ref{fig3}(b)]. The electric-dipole active ($E^\omega \parallel a$) excitation almost disappears above $T_{\rm N}^{\rm Mn}$ [Fig. \ref{fig3}(a)] and its energy range (0--10 meV) nearly coincides with that of the magnon band, which was revealed by inelastic neutron scattering experiments on the $bc$ spiral spin phase of TbMnO$_3$.\cite{RKajimoto,DSenff} This coincidence of the spectral energy region also ensures that the observed excitation is magnetic in origin.

%******************************************************** Figure 6
\begin{figure}[tbh]
\includegraphics[width=0.35\textwidth]{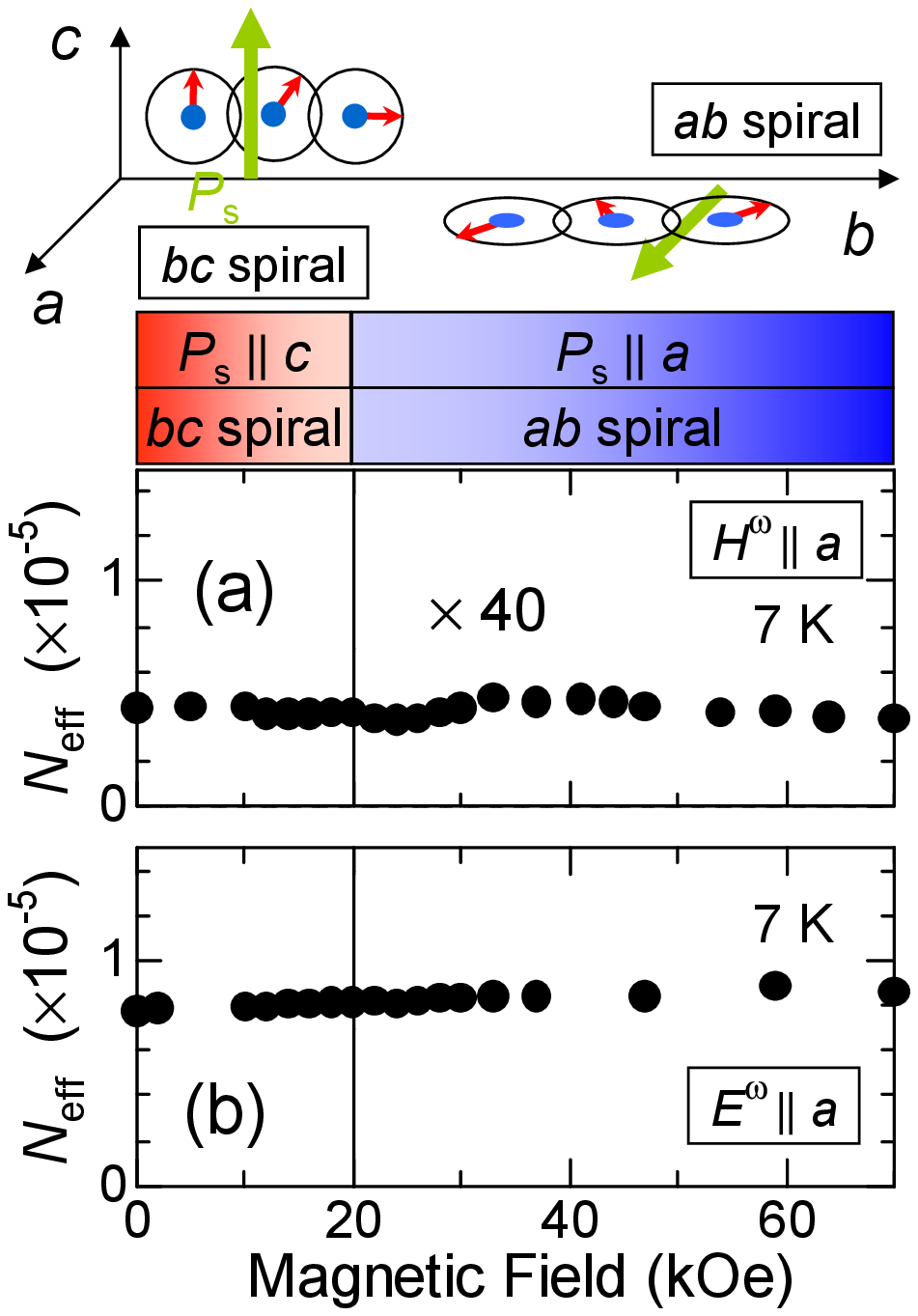}
\caption{(Color online) Magnetic field $(H)$ dependence of the integrated spectral weight per Mn-site, $N_{\rm eff}$, as defined by Eq. (\ref{eqn.Neff}) in the text, of the $ac$ surface crystal plate of DyMnO$_3$ for $H^\omega \parallel a$ [(a)] and $E^\omega \parallel a$ [(b)], measured at 7 K. Upper panel of (a) shows the phase diagram with variation of $H$. We also show the schematic illustrations of the $bc$ and $ab$ spiral spin planes induced by $H$. $H$ of 20 kOe at which the ferroelectric polarization flop occurs at 7 K, is indicated by the vertical solid line in (a) and (b). The difference of $N_{\rm eff}$ at 7 K with [in (a) and (b)] and without $H$ [in Fig. \ref{fig4}], arises from the difference of the integrated energy range (see text). }
\label{fig6}
\end{figure}
%********************************************************

\subsection{Magnetic field dependence}\label{ResultsH}

As an origin of the electromagnon, the rotational mode of the spiral spin plane around the $b$-axis (propagation vector) has been theoretically proposed based on the spin-current model.\cite{HKatsura2} This was taken into consideration in the analysis of the inelastic neutron scattering results for the $bc$ spiral spin phase of TbMnO$_3$\cite{DSenff} and the observed dispersion relation (2.4 meV at $k=0$) was assigned to the rotational mode of the $bc$ spiral spin plane. These apparently contradict the results of $\epsilon$ spectra of GdMnO$_3$,\cite{APimenov1,APimenov2} in which the peak structure with $Im[\epsilon]\sim 1$, comparable to $Im[\epsilon]$ of TbMnO$_3$, was also identified in the IC spin-collinear paraelectric phase. In the course of the $P_{\rm s}$ flop from the $c$ to $a$ axis at $H_{\rm c}$ $\sim$ 20 kOe, IC $q_b^{\rm Mn}$ $(\sim0.37)$ is kept nearly constant for DyMnO$_3$,\cite{JStrempfer} as contrasted by the IC to commensurate transition for TbMnO$_3$.\cite{TArima} Therefore, the measurements of the effect of $H$ on $\epsilon\mu$ spectra of DyMnO$_3$ can provide further insights into the nature of the electromagnon.

The upper and lower panels of Fig. \ref{fig5}(a) present the selected $Re[\epsilon\mu]$ and $Im[\epsilon\mu]$ spectra for $E^\omega \parallel a$, measured at 7 K for varying $H$ with $H\parallel b$-axis, respectively. The experimental setup with use of a superconducting magnet (up to 70 kOe) restricted the measured energy range, and thus we concentrated on the behavior of $\epsilon\mu$ spectrum below 5 meV. Even though the low-lying peak structure grows in intensity and its peak position is shifted to the lower energy above $H_{\rm c}$, the broad continuum-like absorption scarcely depends on $H$ ($H_{\rm c}\approx20$ kOe at 7 K); the $E^\omega\parallel a$ polarized absorption is similarly identified as the most notable electric-dipole active band even in the $ab$ spiral spin phase. Such a tendency is also seen in $H$ dependence of $N_{\rm eff}$, as shown in Fig. \ref{fig6}(b). We selected $\omega_1$ = 0.7 meV and $\omega_2$ = 5 meV to estimate $N_{\rm eff}$ of the low-lying excitation. The negligible effect of $H$ is clearly discerned even $H$ exceeds $H_{\rm c}$. This behavior is different from the reported result of TbMnO$_3$, in which the low-lying electromagnon disappears by application of $H$ up to 80 kOe at 12 K (Ref. \onlinecite{APimenov1}). In the experiment of Ref. \onlinecite{APimenov1}), $H$ was applied along the $c$-axis, thus giving rise to the suppression of $P_{\rm s}$ along the $c$-axis, according to the ME phase diagram of TbMnO$_3$.\cite{TKimura2} Namely, the flop of $P_{\rm s}$ from along the $c$-axis to the $a$-axis or equivalently the rotation of the spiral spin plane from $bc$ to $ab$ cannot be induced unless the $H$ is applied along the $a$ or $b$-axis, as we adopted here. In addition, we measured the $\epsilon\mu$ spectra for $E^\omega\parallel a$ at 17 K just below $T_{\rm c}$ in $H$ along the $b$-axis up to 70 kOe, to confirm the same tendency. Furthermore, we gathered the $\epsilon\mu$ spectra for $E^\omega \parallel c$ and $H^\omega \parallel a$, at 7 K in $H$ up to 70 kOe. The upper and lower panels of Fig. \ref{fig5}(b) shows the selected $Re[\epsilon\mu]$ and $Im[\epsilon\mu]$ spectra, respectively. The estimated $N_{\rm eff}$ is also presented in Fig. \ref{fig6}(a) (we used $\omega_1$ = 2.4 meV and $\omega_2$ = 4 meV). Above $H_{\rm c}$, the direction of $E^\omega$ $(\parallel c)$ becomes perpendicular to the spiral spin plane and hence would activate the electromagnon arising from the rotational mode of the $ab$ spiral spin plane. Even though the evolution of the tiny absorption is visible around 2 meV above $H_{\rm c}$, the strong absorption, to be compared with $Im[\epsilon\mu]$ for $E^\omega \parallel a$ in zero $H$, is not discerned at all for $E^\omega \parallel c$, but only for $E^\omega \parallel a$.

%***********************************************************************
\section{Discussion}\label{Discussion}

In the spiral spin phase of $R$MnO$_3$, the electromagnon or equivalently the rotational mode of the spiral spin plane is expected to become active when $E^\omega$ is set perpendicular to the spiral spin plane, i.e., $E^\omega\parallel a$ and $E^\omega\parallel c$ in $bc$ and $ab$ spiral spin phases, respectively. This scenario is theoretically considered on the basis of the spin-current mechanism (Ref. \onlinecite{HKatsura2}) that is now recognized to explain the magnetically driven ferroelectricity in a variety of multiferroics found so far, including $R$MnO$_3$.\cite{HKatsura1} Indeed, a single peak structure around 2.5 meV has been identified in the $\epsilon$ spectrum for $E^\omega \parallel a$ in the $bc$ spiral spin phase of TbMnO$_3$ (Ref. \onlinecite{APimenov1}). Accordingly, one of low-energy branches of the magnon band, i.e., around 2 meV at $k=0$, in the $bc$ spiral spin phase of TbMnO$_3$ was assigned to be responsible for the electromagnon model by inelastic neutron scattering experiments (Ref. \onlinecite{DSenff}).

In the present study, we could uncover new features of the low-energy spin excitation driven by either $E^\omega$ or $H^\omega$ in a variety of spin phases of DyMnO$_3$ on the basis of the detailed light-polarization, $T$, and $H$ dependences, as presented in Sec. \ref{Results}. First, our result can provide the compelling evidence that the possibility of the rotational mode of the spiral spin plane is excluded as the origin of the intense absorption around 2--3 meV for $E^\omega\parallel a$, that is also observed in the $bc$ spiral spin phase of TbMnO$_3$;\cite{APimenov1} we firmly identify the electric-dipole active spin excitation along the $a$-axis well survives even in the $ab$ spiral spin phase induced by applying $H$ up to 70 kOe along the $b$-axis [Fig. \ref{fig5}(a)]. This observation clearly contradicts the former theoretical\cite{HKatsura2} and experimental\cite{DSenff} arguments described above. Second, we clarify that the electric-dipole active spin excitation along the $a$-axis can be regarded as the continuum-like absorption band with two broad peak structures (1--8 meV) [Figs. \ref{fig2}(a) and \ref{fig2}(b)] rather than the single peak structure (1--5 meV), as reported in the $bc$ spiral spin phase of TbMnO$_3$ (Ref. \onlinecite{APimenov1}). Third, we identify the spin excitation with a single peak structure when $H^\omega$ is set parallel to the $a$-axis or the $c$-axis [Figs. \ref{fig2}(d)--\ref{fig2}(f)], which can be assigned to the antiferromagnetic resonance (AFMR) of Mn spins at $k=0$. The $k=0$ magnon mode for $H^\omega\parallel a$ can be reasonably ascribed to one of the lower branches of the magnon, as observed around 2 meV at $k=0$ in the inelastic neutron scattering spectrum of the $bc$ spiral spin phase of TbMnO$_3$, even though it was previously assigned to the rotational mode of the spiral spin plane.\cite{DSenff} This complexity partly comes form the fact that the peak position of AFMR for $H^\omega\parallel a$ nearly matches the peak position of the lower-lying peak structure for $E^\omega\parallel a$, as we revealed here [Figs. \ref{fig2}(a) and \ref{fig2}(d)]. On the contrary, the observed broad continuum-like band for $E^\omega\parallel a$ spreads over the wide energy range (1--10 meV), which cannot be explained by $k=0$ magnon absorption alone.

%******************************************************** Figure 7
\begin{figure*}[bt]
\includegraphics[width=0.94\textwidth]{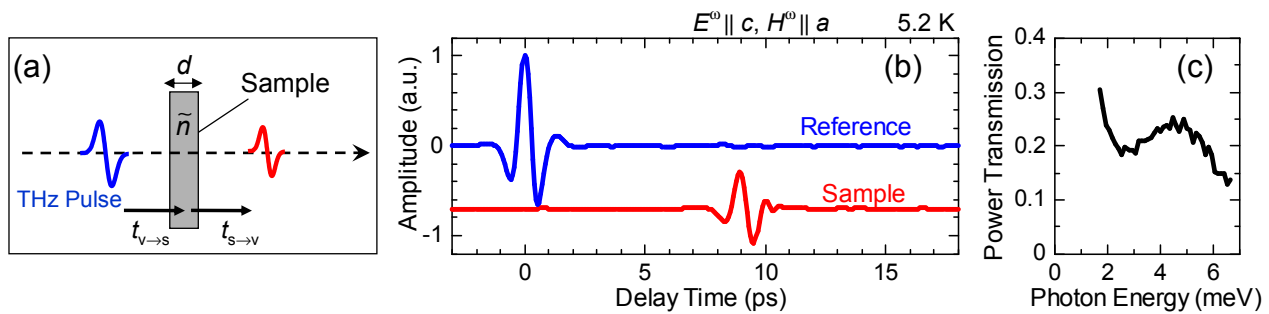}
\caption{(Color online) (a) Sample geometry with the complex Fresnel transmission coefficients from vacuum to the sample $t_{\rm v \rightarrow s}$ and from the sample to vacuum $t_{\rm s \rightarrow v}$. The complex transmission coefficient $t$ was obtained by Eq. (\ref{T}). $d$ and $\tilde{n}$ are the thickness and the complex refractive index of the sample, respectively. (b) Typical example of the measured THz wave form in time-domain with and without the sample ($ac$ surface crystal plate of DyMnO$_3$), measured at 5 K in zero magnetic field. $E^\omega$ and $H^\omega$ were set parallel to the $c$ and $a$-axes, respectively. The amplitude of the transmitted THz wave form with the sample was vertically offset for clarity. (c) Power transmission spectrum of DyMnO$_3$.}
\label{fig7}
\end{figure*}
%********************************************************

The observed broad peak and continuum-like (0--10 meV) shape of the electric-dipole active band does suggest that the spin fluctuation related to the excitation such as the two-magnon is more plausible rather than the Goldstone mode like electromagnon. The optically active two-magnon excitations are possible over the full magnon bandwidth with the combination of $+k$ and $-k$ magnons, and hence should distribute the spectral weight up to the energy twice of the magnon bandwidth. According to the inelastic neutron scattering study on TbMnO$_3$ (Refs. \onlinecite{RKajimoto,DSenff}), the magnon band width is about 8 meV. This value is considerably narrowed as compared with that of LaMnO$_3$ ($\sim32$ meV)\cite{KHirota} or PrMnO$_3$ ($\sim20$ meV)\cite{RKajimoto} with the $A$-type antiferromagnetic ground state, due to the larger orthorhombic lattice distortion as well as to the strong spin frustration effect in the cycloidal magnet TbMnO$_3$. The magnon bandwidth of DyMnO$_3$, in which the inelastic neutron scattering study would be difficult to perform due to the presence of the neutron-absorbing Dy ions, should be comparable to, or slightly smaller than, that of TbMnO$_3$, since DyMnO$_3$ is structurally close to TbMnO$_3$ and shows the similar $bc$ plane cycloidal order at the ground state.\cite{TKimura2} Therefore, two-magnon energy range is anticipated to be up to 10--15 meV, which well fits the observed energy range of the electric-dipole active spin excitations.

Here we propose that the observed unique optical dipole activity may share the same origin, i.e., the spin-current mechanism,\cite{HKatsura1} with the static $P_{\rm s}$. The strong $E^\omega \parallel a$ selection rule irrespective of the $bc$ or $ab$ spiral spin plane does suggest the large mutual fluctuation within the $ac$ plane, of the neighboring spins lining up along the $c$-axis, but not along the $b$ nor $a$-axis; namely $(\delta S_i \times \delta S_j) \parallel b$, $e_{ij} \parallel c$, and hence $\delta P^\omega$ $[\propto e_{ij}\times(\delta S_i \times \delta S_j)] \parallel a$. The antiferromagnetic exchange interaction along the $c$-axis in the perovskite manganite is likely the origin of the large spin fluctuation, contrary to the ferromagnetic coupling along the $a$-axis. This scenario, though a more elaborate theory is needed, may qualitatively explain the strong optical activity for $E^\omega \parallel a$ also in the spin-collinear paraelectric phase as observed. Then, broad peaks around 2 meV and 5.5 meV may be assigned to the density-of-state singularities of the two-magnon band.

\section{Summary}%***********************************************
\label{Summary}

In summary, we have measured the $\epsilon\mu$ spectra of a complete set of possible configurations of single crystals of DyMnO$_3$ by using THz time-domain spectroscopy. A rich variety of optical excitation at THz frequencies was identified by detailed measurements of light-polarization, $T$, and $H$ dependences. In particular, the spin excitation driven by $E^\omega$ was confirmed to become strongly electric-dipole active along the $a$-axis, irrespective of the $bc$ or $ab$ spiral spin plane. The broad (1--10 meV) band feature indicates the contribution of two-magnon like excitations arising from the large fluctuations of spin and spin-current inherent to this spiral magnet.

%***********************************************
\begin{acknowledgments}
We thank F. Kagawa for characterization of the samples used in this study by measurements of $\epsilon$ at low-frequencies ($\sim$ 10 kHz) and S. Miyahara, M. Mochizuki, N. Furukawa, H. Katsura, and Y. Taguchi for enlightening discussions. This work was in part supported by Grant-In-Aids for Scientific Research (16076205 and 20340086) from the Ministry of Education, Culture, Sports, Science and Technology (MEXT), Japan.
\end{acknowledgments}

%******************************************************** Figure 8
\begin{figure}[bt]
\includegraphics[width=0.48\textwidth]{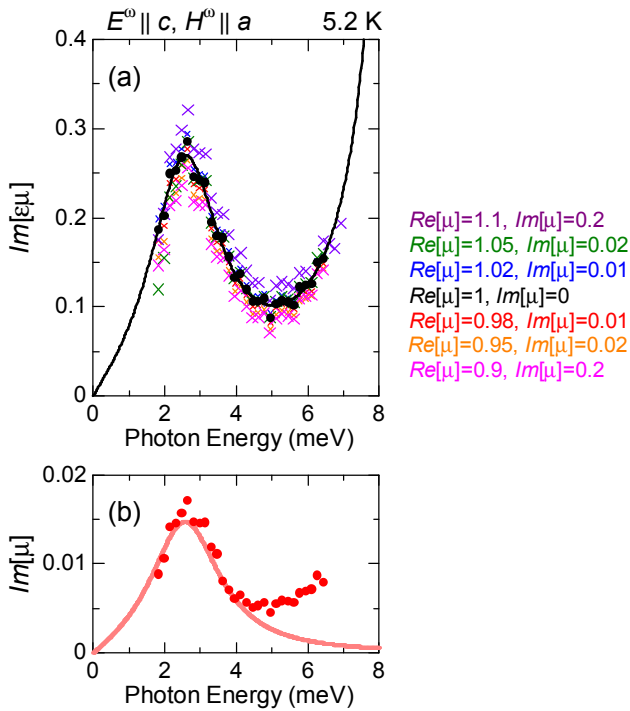}
\caption{(Color online) (a) Validity of using Eqs. (\ref{trans1}) and (\ref{trans2}), instead of Eqs. (\ref{trans1m}) and (\ref{trans2m}). The $Im[\epsilon\mu]$ spectra of DyMnO$_3$ for $E^\omega \parallel c$ and $H^\omega \parallel a$ at 5 K was almost unchanged even if we used Eqs. (\ref{trans1m}) and (\ref{trans2m}). The fixed parameters of $Re[\mu]$ and $Im[\mu]$ are described in the right-hand side of the figure, which were confirmed to be reasonable by assuming that $\epsilon$ can be represented by a single Lorentz oscillator [see, (b)]. (b) The estimated $Im[\mu]$ spectrum (closed circles) by assuming that $\epsilon$ can be represented by a single Lorentz oscillator. We also plot the $Im[\mu]$ spectrum (a solid line), as extracted from Eq. (\ref{T}) combined with Eqs. (\ref{trans1}) and (\ref{trans2}).}
\label{fig8}
\end{figure}
%********************************************************

%*********************************************************************
\appendix
\section{Raw data of terahertz radiation in time-domain}

A unique advantage for the use of THz time-domain spectroscopy is to directly monitor the electric field of the radiated THz pulse in time-domain.\cite{DGrischkowsky} This technique yields the amplitude and phase spectra by the fast Fourier transformation (FFT) of the measured THz wave forms with and without the sample. Therefore, we need no Kramers-Kronig transformation, which is necessary for obtaining the complex optical spectra by conventional optical spectroscopy. As a typical example, we show in Fig. \ref{fig7}(b) the transmitted THz wave form with the sample in time-domain, performed at 5.2 K in zero $H$. The sample was the $ac$ surface crystal plate of DyMnO$_3$. $E^\omega$ and $H^\omega$ were set parallel to the $c$ and $a$-axes, respectively. We also plotted the reference of THz pulse used in this setup. The amplitude of the transmitted THz wave form is reduced due to the absorption of the sample, as compared to the reference. Its phase is also delayed due to the refractive index of the sample. The ratio between FFT spectrum with and without the sample, 
\begin{equation}
\frac{E_{\rm sample}}{E_{\rm ref}}\equiv\sqrt{\mathcal{T}}\exp({-i\phi})
\label{Meas}
\end{equation}
can yield the power transmission $\mathcal{T}$ and phase $\phi$ spectra. The example of the obtained $\mathcal{T}$ spectrum is shown in Fig. \ref{fig7}(c). We can identify the clear absorption peak structure around 3 meV. This peak can be assigned to the spin excitation driven by $H^\omega \parallel a$, as presented in Fig. \ref{fig2}.

\section{Relationship between transmission coefficient and optical constants}

We numerically derived the complex optical spectra without the KK transformation by adopting the following procedure. Without taking into account the effect of the multiple reflections inside the sample, the complex transmission coefficient $t$ for bulk in normal incidence as schematically shown in Fig. \ref{fig7}(a), can be simply expressed as
\begin{equation}
t=t_{\rm v \rightarrow s}t_{\rm s \rightarrow v}\exp\left(-i\frac{\omega}{c}d(\tilde{n}-1)\right)=\frac{E_{\rm sample}}{E_{\rm ref}},
\label{T}
\end{equation}
where $\omega$ is the frequency, $c$ the velocity of light, $d$ the thickness of the sample, and $\tilde{n}$ $(=n+i\kappa)$ the complex refractive index of the sample.\cite{BornWolf} Experimentally, the effect of the multiple reflections can be avoided by restricting the time range of the FFT. The complex Fresnel transmission coefficients\cite{BornWolf} from vacuum to the sample $t_{\rm v \rightarrow s}$ and from the sample to vacuum $t_{\rm s \rightarrow v}$, can be, respectively, given by
\begin{subequations}
\begin{eqnarray}
t_{\rm v \rightarrow s}&=&\frac{2}{\tilde{n}+1},\label{trans1}\\
t_{\rm s \rightarrow v}&=&\frac{2\tilde{n}}{\tilde{n}+1},
\label{trans2}
\end{eqnarray}
\end{subequations}
when the complex magnetic permeability $\mu$ is 1. In this case, $\tilde{n}$ can be expressed as $\sqrt{\epsilon}$, where $\epsilon$ is the complex dielectric constant of the sample. Accordingly, $\tilde{n}$ and thus $\epsilon$ can be numerically calculated from Eq. (\ref{T}) using experimentally determined values; $\mathcal{T}$ and $\phi$ in Eq. (\ref{Meas}). This procedure is widely recognized to estimate $\tilde{n}$ and $\epsilon$ of a variety of materials including dielectrics, semiconductors, and metals.\cite{THzRev,THzRev1}

On the contrary, in magnetic materials, $\tilde{n}$ should be exactly expressed as $\sqrt{\epsilon\mu}$ near the magnetic resonance, which usually locates around $\sim$ meV (Ref. \onlinecite{AMBalbashov}). In this case, Eqs. (\ref{trans1}) and (\ref{trans2}) should be, respectively, replaced by 
\begin{subequations}
\begin{eqnarray}
t_{\rm v \rightarrow s}&=&\frac{2\mu}{\tilde{n}+\mu},\label{trans1m}\\
t_{\rm s \rightarrow v}&=&\frac{2\tilde{n}}{\tilde{n}+\mu}.
\label{trans2m}
\end{eqnarray}
\end{subequations}

\section{Estimation procedure of complex optical spectra}

In the cases of the measurements for $E^\omega \parallel a$ using $ac$ and $ab$ surface crystal plates of DyMnO$_3$, we experimentally identified from light-polarization dependences [Figs. \ref{fig2}(a) and \ref{fig2}(b)] that the observed large absorption originates from $E^\omega$ response. The $H^{\rm \omega}$ response in these configurations can be considered to be negligible compared to $E^\omega$ response. This is confirmed from the results using $bc$ surface crystal plate where there is no absorption in $H^\omega\parallel b$ ($E^\omega\parallel c$) [Fig. \ref{fig2}(c)], and only a slight absorption in $H^\omega\parallel c$ ($E^\omega\parallel b$) [Fig. \ref{fig2}(f)]. Therefore, we assume $\mu=1$ in the Fresnel coefficient and used Eqs. (\ref{trans1}) and (\ref{trans2}), while keeping $n=\sqrt{\epsilon\mu}$ in Eq. (\ref{T}) for generality. This is also applied to the case for $E^\omega \parallel c$ and $H^\omega \parallel b$, in which the negligible absorption is clearly identified [Fig. \ref{fig2}(c)]. We numerically solved Eq. (\ref{T}) combined with Eqs. (\ref{trans1}) and (\ref{trans2}) to match the measured values given by Eq. (\ref{Meas}) and determined the $\epsilon\mu$ spectra. For the sake of convenience, we used the quantities of $Re[\epsilon\mu]$ and $Im[\epsilon\mu]$ in this paper, although they can be precisely expressed as $Re[\epsilon]$ and $Im[\epsilon]$ in the above condition.

On the other hand, we identify a peak structure around 3 meV, which only appears when $H^\omega$ was set parallel to the $a$-axis [Figs. \ref{fig2}(d) and \ref{fig2}(e)] or the $c$-axis [Fig. \ref{fig2}(f)]. These small absorption can be assigned to $H^\omega$ response. In addition, a broad background absorption is observed above 4 meV as seen in the power transmission spectrum at 5.2 K [Fig. \ref{fig7}(b)], which can be attributed to the tail of the optical phonon lying around 16 meV.\cite{APimenov2} In such a case, both $\epsilon$ and $\mu$ could contribute to the transmission spectrum. Accordingly, four independent quantities, $Re[\epsilon]$, $Im[\epsilon]$, $Re[\mu]$, and $Im[\mu]$, cannot be obtained only from the transmission measurements of $\mathcal{T}$ and $\phi$. However, in the case of a thick sample, the main signature of the transmission spectrum is dominated by bulk absorption and the effect of $\mu$ in the Fresnel coefficient is considered to be small, as discussed below. Therefore, we used the Fresnel coefficient of Eqs. (\ref{trans1}) and (\ref{trans2}), while the effect of $H^\omega$ response was taken into account by adopting $\tilde{n}=\sqrt{\epsilon\mu}$ in Eq. (\ref{T}). The validity of this approach was firmly confirmed following calculation.

We numerically solved Eq. (\ref{T}) by putting Eqs. (\ref{trans1m}) and (\ref{trans2m}) with $Re[\mu]$ and $Im[\mu]$ as fixed parameters with respect to $\omega$ and estimated $Im[\epsilon\mu]$ spectra of DyMnO$_3$ for $E^\omega \parallel c$ and $H^\omega \parallel a$ at 5.2 K. For examples, we show in Fig. \ref{fig8}(a) the calculated $Im[\epsilon\mu]$ spectra at selected values of $Re[\mu]$ and $Im[\mu]$. The fixed parameters used here are provided in the right-hand side of the figure. The $Im[\epsilon\mu]$ spectrum, as derived from Eqs. (\ref{trans1}) and (\ref{trans2}) [$Re[\mu]$ = 1 and $Im[\mu]$ = 0 in Eqs. (\ref{trans1m}) and (\ref{trans2m})], is also plotted by closed circles with a solid line. As can be clearly seen in the figure, no significant difference is observed even though $\mu$ is included to $t_{\rm v \rightarrow s}$ and $t_{\rm s \rightarrow v}$.

To further check the validity of the used values of $Re[\mu]$ and $Im[\mu]$, we assumed that $\epsilon$ comes from the contribution of the optical phonon structure, which can be represented by a single Lorentz oscillator. This approximation has been conventionally used to estimate the $\mu$ spectra in antiferromagnets, as exemplified by the cases of the orthoferrites.\cite{AMBalbashov} We used the following parameters; the resonance frequency, the damping rate, and the dielectric constant for the high frequency above the measured range, were set to 9 meV, 0.5 meV, and 16, respectively. With this approximation, the $Im[\mu]$ spectrum was numerically calculated using Eq. (\ref{T}) combined with Eqs. (\ref{trans1m}) and (\ref{trans2m}). The result is presented in Fig. \ref{fig8}(b) by closed circles. We also show the $Im[\mu]$ spectrum (a solid line), as derived from the $Im[\epsilon\mu]$ spectrum that was estimated by using Eqs. (\ref{trans1}) and (\ref{trans2}). We assumed that the $Im[\epsilon\mu]$ spectrum $(=Re[\epsilon]Im[\mu]+Im[\epsilon]Re[\mu])$ can be expressed by two Lorentz oscillators for $\epsilon$ and $\mu$. The both data coincide with each other below 4 meV, ensuring the validity of using Eqs. (\ref{trans1}) and (\ref{trans2}). Noticeably, the maximum values of $Re[\mu]$ and $Im[\mu]$ are estimated to be about 1.01 and 0.02, respectively, indicating that the fixed parameters shown in Fig. \ref{fig8}(a) are reasonable. However, there is a discrepancy above 4 meV, which mainly arises from the lack of the spectral information above 6 meV and thus from the inappropriate parameters used here to reproduce the optical phonon structure. In this work, we numerically solved using Eq. (\ref{T}) with Eqs. (\ref{trans1}) and (\ref{trans2}) and estimated $\tilde{n}$ $(=\sqrt{\epsilon\mu})$ spectra.

%****************************************************************

\end{document}